\documentclass{article}
\usepackage{graphicx}
\usepackage{authblk}
\begin{document}
\title{Big Data: the End of the Scientific Method?}
\author[1,2]{Sauro Succi}
\author[3,4]{Peter V. Coveney\footnote{Corresponding author: Peter V. Coveney, p.v.coveney@ucl.ac.uk}}
\affil[1]{Center for Life Nano Sciences at La Sapienza,\\ Istituto Italiano di Tecnologia, viale R. Margherita, Roma, Italy}
\affil[2]{Institute for Applied  Computational Science,\\ J. Paulson School of Engineering and Applied Sciences,\\ Harvard University, 33 Oxford Street, Cambridge, USA}
\affil[3]{Centre for Computational Science, Department of Chemistry, University College London}
\affil[4]{Yale University, USA}

\maketitle

\begin{abstract}
We argue that the boldest claims of Big Data are in need of revision and toning-down, in view of a few basic lessons learned from the science of complex systems. We point out that, once the most extravagant claims of Big Data are properly discarded, a synergistic merging of BD with big theory offers considerable potential to spawn a new scientific paradigm capable of overcoming some of the major barriers confronted by the modern scientific method originating with Galileo. These obstacles are due to the presence of nonlinearity, nonlocality and hyperdimensions which one encounters frequently in multiscale modelling.
\end{abstract}

\section{Introduction}
Our current society is characterised by an unprecedented ability to produce and store breathtaking amounts of data and, much more importantly, by the ability to navigate across them in such a way as to distil useful information, hence knowledge, therefrom. This has now reached the point of spawning a separate discipline, so-called ``Big Data'' (BD), which has taken the scientific and business domains by storm. Like all technological revolutions, the import of BD goes far beyond the scientific realm, reaching down into deep philosophical and epistemological questions, not to mention societal ones. One of most relevant is: are we facing a new epoch in which the power of data renders obsolete the use of the scientific method as we have known it since Galileo?
\noindent
That is, insight gained through a self-reinforcing loop between experimental data and theoretical analysis, based on the use of mathematics and modelling?

For if, as sometimes appears true today, anything can be inferred by detecting patterns within huge databases, what's the point of modelling anymore? This extreme stance is summarised in Anderson's provocative statement: ``With enough data, the numbers speak for themselves, correlation replaces causation, and science can advance even without  coherent models or unified theories''. In a nutshell, it is a data-driven version of Archimedes' fulcrum: give me enough data and I shall move the world.
As radical as this new empiricism is, it brings up an intriguing point: is understanding overrated? Could it be that smart algorithmic search through oceans of data can spare us the labour (and the joys) of  learning how the world works\cite{STROGA}? 

``Why learn, if you can look it up''?  is another way of articulating the same idea. At least among intellectuals, the retort is that of C.S. Lewis, ``Once you have surrendered your brain, you've surrendered your life'' (paraphrased) \cite{CSL}. In the sequel, we shall offer rational arguments in support of this instinctive reaction whilst recognising the perspectives opened up by BD approaches.
 
\section{Why is Big Data so sexy?}

Big Data flourishes upon four main observations, namely:

(i) The explosive growth of data production/acquisition/navigation capabilities.

(ii) Reading off patterns from complex datasets through smart search algorithms may be faster and more revealing  than modelling the underlying behaviour i.e. using theory.

(iii) It applies to {\it any} discipline, including those traditionally not deemed suitable for mathematical treatment, including Life Sciences (another way of putting this is to suggest that these domains are too complex to be modelled).

(iv) Its involvement in immediate application to business and politics, ``opinion dynamics'', ``sentiment analysis'' and so on, furnishes another set of domains which raise many ethical questions. 

While the four points above hold disruptive potential for science and society, in the following we shall illustrate how and why, based on basic findings within the modern science of complexity, all of them may lead to false expectations and, at their nadir, even to dangerous social, economical and political manipulation.  

The four points we shall make in response are the following:

\begin{enumerate}
\it{
\item Complex systems are strongly correlated, hence they do not (generally) obey Gaussian statistics.
\item No data is big enough for systems with strong sensitivity to data inaccuracies.
\item Correlation does not imply causation, the link between the two becoming exponentially fainter at increasing data size.
\item In a finite-capacity world, too much data is just as bad as no data.
}
\end{enumerate}
Far from being exceptional, our four assertions apply to most complex systems of relevance to modern science and society, such as far from equilibrium nonlinear physics, finance, wealth distribution and many social phenomena as well. So, there can be no excuse for ignoring them.
 
\section{Complex systems do not (generally) obey Gaussian statistics}

Big Data radicalism draws heavily upon a fairly general fact of life: the Law of Large Numbers (LLN), the main content of which is that, with enough samples, call it $N$, error (uncertainty) are bound to surrender to certainty. The most famous aspect of which is the square-root law of the noise/signal ratio:
\begin{equation}
\label{LLN}
\frac{\sigma}{m} \propto \frac{1}{\sqrt N}
\end{equation}
where $m$ is the mean value and $\sigma$ its root-mean-square associated to a given stochastic process $X$. In other words, let $m_N = (x_1 + \dots x_N)/N$ be the mean value of a given quantity $X$ as obtained from $N$ measurements i.e. the data. It is well known that $m_N$ approaches the correct mean, $m$, in the limit of $N \to \infty$. Even better, one can estimate how fast such convergence is attained by inspecting the mean square departure from the mean, also known as the variance, namely:  
\begin{equation}
\sigma_N^2 = \frac{1}{N} \sum_{i=1}^N (x_i-m_N)^2, 
\end{equation}
Under fairly general assumptions, it can be shown that the root-mean-square (rms) departure from the mean decays like $1/\sqrt{N}$. With enough measurements, uncertainty surrenders: this is the triumph of Big Data \cite{S2N}.

Now let us ask ourselves: what are the ``general assumptions'' we alluded to above? The answer is that the variables $x_i$ must: (i) be uncorrelated, i.e. each outcome $x_i$ is independent of the previous one and does not affect the next one either, (ii) exhibit a finite variance. As we shall see, neither of the two should be taken for granted.

With these two premises, the central limit theorem pertaining to LLN shows that the sum $X_N$ obeys Gaussian statistics, i.e. a bell shaped curve
\begin{equation}
\label{GAUSS}
p_G(y) = \frac{1}{2 \pi \sigma} e^{-y^2/2}
\end{equation}
where $y=(x-m)/\sigma$ is the normalized (de-trended) version of $x$ (here $x$ stands for any generic stochastic variable).

The Gaussian distribution exhibits many important properties, but here we shall focus on the following one: {\it Outliers stand very poor chances of manifesting themselves, and precisely because they are the carriers of uncertainty, uncertainty is heavily suppressed}. Then the numbers indeed speak for themselves: the probability of finding an event one-sigma away from the mean is about $30$ percent, a number which goes down to just $4.5$ percent at two-sigma. The demise of uncertainty is dramatic, at five-sigma, we find just about half a million and at six-sigma less than two in a billion! This adumbrates a very comfortable world, where uncertainty has no chances because outliers are heavily suppressed; fluctuations recede and are absorbed within the mean, an overly powerful attractor. A comfortable, if somewhat grey, world of stable and reassuring conformity.  

The Gaussian distribution plays a undeniable role across all walks of science and society, to the point of still being regarded by many as a universal descriptor of uncertainty. The truth is that, for all its monumental importance, the Gaussian distribution is far from being universal. In fact, it fails to describe most phenomena where complexity holds sway.

{\it Why?} Basically, because: complex systems, almost by definition, are correlated! When a turbulent whiff is ejected from the wind-shield of our car, it affects the surrounding air flow, so that the next whiff will meet with an environment which is not the same it would have met in the absence of the previous whiff. The system affects its environment, the two are correlated, the statistics of whiffs (turbulence) is not Gaussian. This is a far cry from the ``fair coin'', in which head or tail now has no effect on head or tail at the next toss. In complex systems the coin is hardly fair. As a result, the statistics of correlated events is much more tolerant towards outliers, with  the consequence of a much higher, sometimes even unbounded, variance. In a nutshell, it is a world much more full of (good and bad) surprises, just as is real life!
\begin{figure}
\centering
\includegraphics[width=0.8\textwidth]{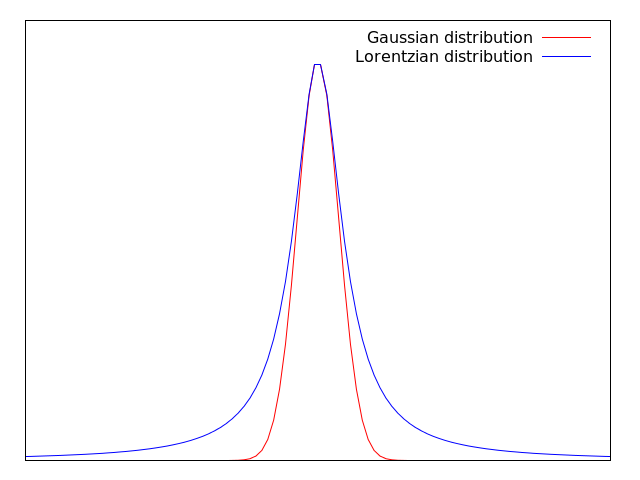}
\caption{A comparison of Gaussian and Lorentzian distributions. Note the persistence in the latter distribution of far larger events from the mean.}
\label{fig:GaussvsLorentz}
\end{figure}

The prototypical example is the Lorentz distribution:
\begin{equation}
\label{LOR}
p_L(x) = \frac{a/\pi}{a^2+(x-m)^2}
\end{equation}

For inliers, $|x-m|<<a$, this is virtually indistinguishable from a Gaussian, and all is fine and well. But for events far in excess of $a$, outliers or {\it rare events} in the following, the difference is dramatic: the Lorentz distribution decays much more slowly than the Gaussian, which is the reason why its variance is formally infinite; see Figure \ref{fig:GaussvsLorentz}. As an example, if human heights were distributed according to Lorentz, with an average $m=1.75$ (meters) and $a=10$ cm, the probability of finding a human $2.75$ meters high, i.e. at $10 \sigma$, would be of the order of three percent. The decay is so slow as to sound ridiculous: going to $20 \sigma$ just halves the number, which means that walking along the street of any city in the world, humans $3.75$ m high would be commonplace!\footnote{For the detail-thirsty, this probability is computed by means of the cumulative distribution of the Lorentz function, namely $P_L(x) = 1/2 + \frac{1}{\pi} atan(\frac{x-m}{a})$.} Ridiculous as it seems for human heights (which are indeed Gaussian-distributed), such slow decay of outliers is a regular occurrence in complex systems, be they natural, financial or social. The dire consequences of treating the financial world as a Gaussian-regulated one are most compellingly (and often hilariously) discussed in Taleb's books ``{\it Fooled by Randomness}'' \cite{TALEB1} and ``{\it The Black Swan}'' \cite{TALEB2}.

Infinite variance is a bit far-fetched, since in the real world signals and measurements are necessarily finite, but the message comes across loud and clear: the mean and the variance are no longer sufficient to capture the statistical nature of the phenomenon. For this purpose higher order {\it moments} must be inspected.

By moments, we mean sums (integrals in continuum space, with due care)  of the form
\begin{equation}
M_q = \sum_{i=1}^N x_i^q 
\end{equation}
where $q$ is usually (but not necessarily) a positive integer. By normalising with $M_0=N$, it is clear that $M_1$ is the mean and $M_2-M_1^2$ is the variance. For a Gaussian distribution, this is all we need to know, because all higher-order  moments with $q>2$ follow directly from these two. But for a generic distribution this is no longer the case and more moments need be specified; in particular, from the very definition, it is readily appreciated that large $q's$ give increasing weight to large $x's$, i.e. the aforementioned rare-events play an increasing role. Thus, inspection of $M_q$ with $q>2$ is paramount to the understanding of complex processes, an utterly non-Gaussian world trailblazed by the turbulence community and now widespread to most walks of the science of complexity. The bottomline is that, in the presence of correlations, Gaussian statistics no longer hold, and uncertainty does not give in so easily under data pressure, in that the convergence to zero uncertainty is much slower than the inverse square-root law. For a moment of order $q$, it is likely to be $N^{-1/q}$, which is a nearly flat in practice for $q>>2$. For instance, with $q=8$, cutting down uncertainty by a factor of $2$ takes $2^8=256$ times more data. 

This explains why the BD trumpets should be toned down: when rare events are not so rare, convergence rates can be frustratingly slow even in the face of petabytes of data. 

\subsection{To Gauss or not to Gauss: nonlinear correlations}

It is natural to ask if there is a qualitative criterion to predict whether a given system would or would not obey Gaussian statistics. While these authors are not aware of any rigorous ``proof'' in this direction, robust heuristics are certainly available. We have mentioned before that the law of large numbers rests on the assumption that the sequence of stochastic events be uncorrelated, that is, the occurrence of a given realisation does not depend on the previous occurrences and does not affect the subsequent ones. This is obviously an idealisation, but one which eminently works well, as long as the system in point can be treated as isolated from its environment and not subject to any form of nonlinearity. Yet, by definition, most complex systems do interact with their environment and they affect it in various ways. Since the environment couples back to the system, it is clear that self-reinforcing or self-destroying loops get set up in the process. Self-reinforcing loops imply that a given occurrence affects the environment in such a way as to make such an occurrence more likely to happen again in the future. This is the basic mechanism giving rise to persistent correlations, the unfair coin we alluded to earlier on in this paper. And persistent correlations are a commonplace in most complex systems, be they natural, financial, political, psychological or social.

\section{Sensitivity to data inaccuracies}
The main goal of BD is to extract patterns from data, i.e. to unveil correlations between apparently
disconnected phenomena.
Given two processes, say $X=\lbrace x_i \rbrace$ and $Y=\lbrace y_i \rbrace$, $i=1,N$
the standard measure of their correlation 
is the {\it covariance}, defined as
\begin{equation}
\label{CORRE}
C(X,Y) = \frac{\sum_{i=1}^N  x_i y_i}{\sigma_x \sigma_y}
\end{equation}
where the sequences $x_i$ and $y_i$ are assumed to be de-trended, i.e. of zero-mean.
The most perfect correlation is $Y=X$, which delivers $C=1$,
its opposite being perfect anti-correlation, $Y=-X$, yielding $C=-1$.
Also interesting is the case of plain indifference, zero-correlation $C=0$, which means that
sets of positively and negatively correlated events are in perfect balance.
\begin{figure}
\centering
\includegraphics[width=0.8\textwidth]{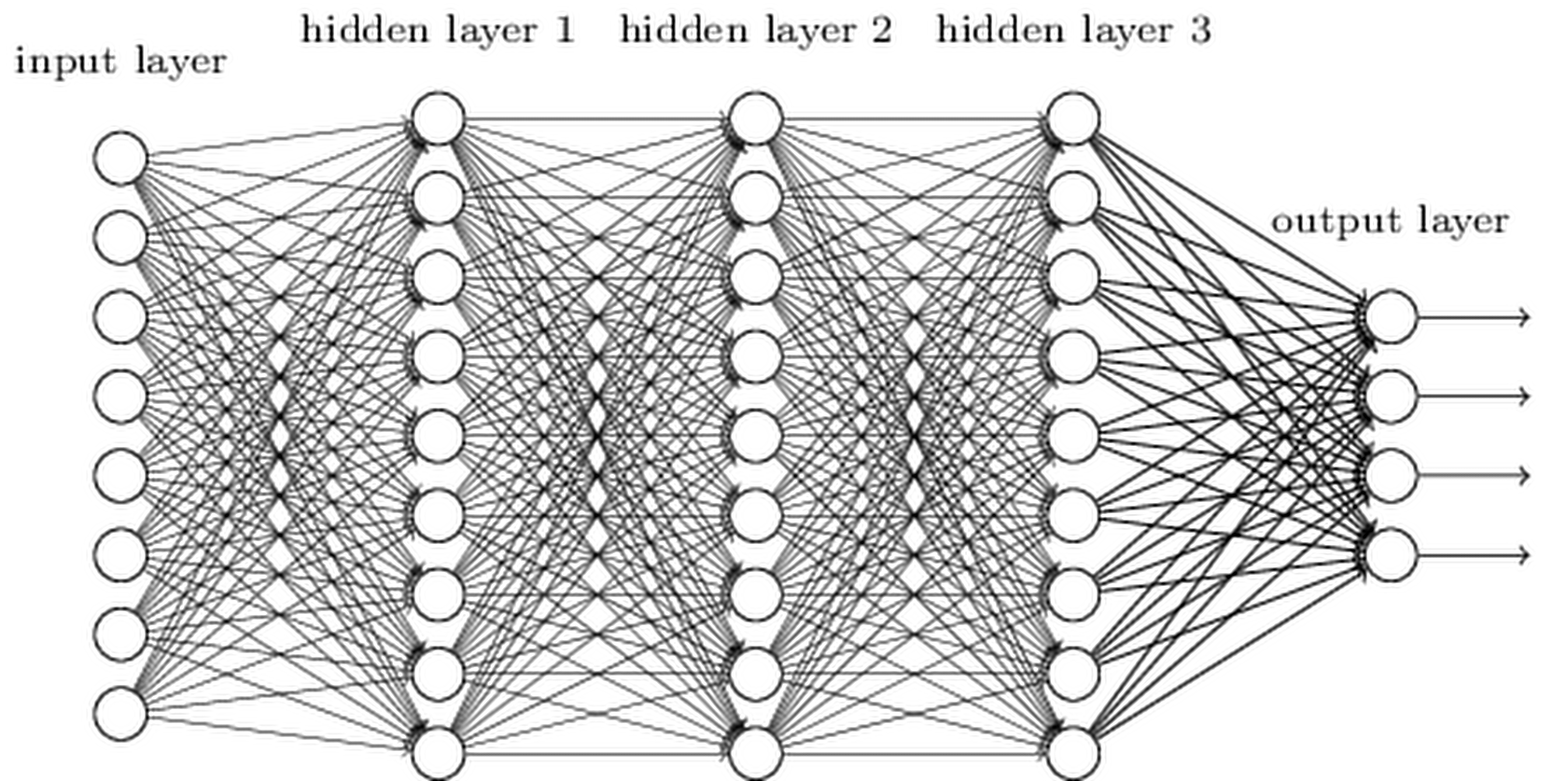}
\caption{Note the presence of multiple hidden layers within which specific processing activity takes place, between the input and output layers of processing elements, called neurons.}
\label{fig:Deepneural}
\end{figure}
In geometrical language, $C=0$ implies that the two N-dimensional  vectors $X$ and $Y$
are orthogonal. Adhering to this language, the correlation $C$ can be thought of as the
cosine of the angle between the vectors $X$ and $Y$, in which case we write
$C = (X,Y)/(\sigma_x \sigma_y)$, where $(,)$ denotes the scalar product in Euclidean space.
Needless to say, under Gaussian statistics, correlation coefficients converge to 
the ``exact'' values in the limit of infinite $N$.

But again, this is not necessarily the case if $X$ and $Y$ stem from complex processes. Moreover, quadratic correlators, such as the covariance in eq. (\ref{CORRE}), which originate directly from the notion
of Euclidean distance $d(X,Y) = \sum_i (y_i-x_i)^2$, may not be adequate to capture the complex
nature of the phenomena, just as mean and variance are 
by no means the full story in the presence of rare events!
In particular, higher order ``distances'', possibly not even Euclidean ones, should be inspected, their
resilience to data pressure being inevitably much higher than the one offered by the variance. 
Once again, error convergence might be a very slow function of data size.

A similar argument goes for more sophisticated forms of learning, such as the currently all-popular {\it machine learning}. Here, a neural net is trained to recognise patterns within a given set of data, by adjusting the weights
of the connections in such a way as to minimise a given error functional (cost function in machine-learning jargon).
Given a set of input data ${x_i}$, the neural net produces a corresponding output ${y_i}$ of the form
$$
y_i = f(\sum_j W_{ij} x_j)
$$ 
where $W_{ij}$ is the connecting weight between nodes $i$ and $j$ belonging to two
subsequent layers of the network and $f()$ a suitable transfer function, typically 
a sigmoid or variations thereof. See Figure \ref{fig:Deepneural}.

The output signal is then compared to the target data $Y_i$ to form the loss function
$$
E \lbrace W \rbrace = \sum_{i=1}^N d(y_i,Y_i)
$$
where $d(x,y)$ is some metric distance in data space, usually, but not necessarily, the standard Euclidean one.

\begin{figure}
\centering
\includegraphics[width=0.6\textwidth]{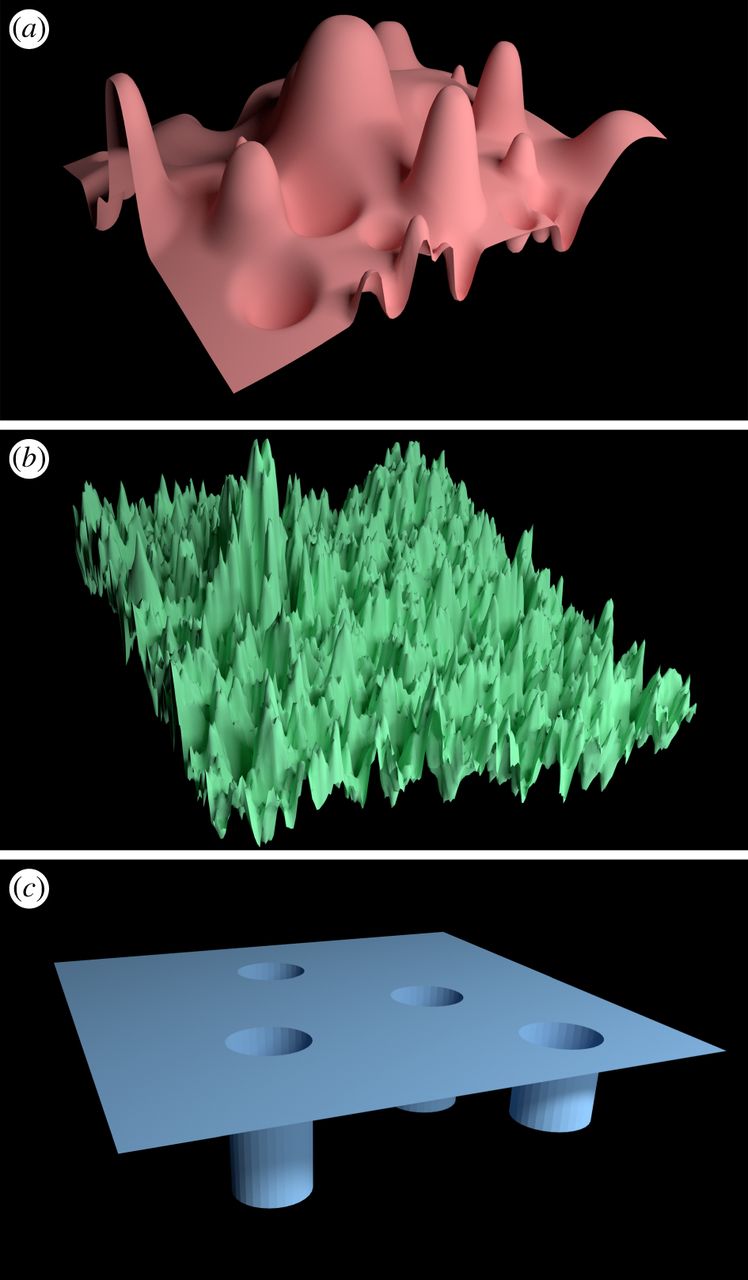}
\caption{Examples of three classes of landscapes that may be encountered in connecting input to output variables, shown in three dimensions but typically in fact arising in much higher dimensions which cannot be drawn. (a) shows a relatively smooth (i.e. continuous) landscape over which machine learning algorithms might be expected to perform well; (b) is a fractal landscape which is not differentiable and contains structure on all length scales; (c) shows another discontinuous landscape with no gradients. Both (b) and c) would not be expected to perform reliably in the context of machine-learning algorithms. Reproduced from Coveney {\em et al.} \cite{PVC1}}
\label{fig:landscapes}
\end{figure}

The weights are then updated according to some dynamic minimisation schedule, so as to achieve the minimum error.
It is then clear that if the functional $E \lbrace W \rbrace$ is smooth, the search is easy and robust against data
inaccuracies. If, on the other hand, the error landscape is corrugated, the expected case for 
complex systems in which higher order moments carry most of the relevant information, even 
small inaccuracies can result in the wrong set of weights, where wrong means that such weights are likely 
to fail when applied to new data beyond those they had been trained for. See Figure \ref{fig:landscapes}

Of course, failsafe scenarios also exist, whereby different sets of weights work even though 
they differ considerably from each other, because all local minima are basically equivalent quasi-solutions
of the optimisation problem. We suspect, without proof, that such a form of benevolence lies at the basis
of most remarkable successes of machine learning and modern artificial intelligence applications.

But this cannot be assumed to be the universal rule: a big red light is there in general, the name of the game 
being ``overfitting'', that is a stiff solution exists which reproduces very well a given set of data, but fails grossly as soon 
as the dataset is enlarged, if only slightly. Although well-known to the machine-learning scientific community, these problems are 
typically swept under the carpet by the most ardent Big Data aficionados.

\section{The two distant sisters: Correlation and Causation}
The fact that correlation does not imply causation is 
such a well-known topic that we only mention it for completeness.

It is indeed well recognised that even if two signals manage to register a very high correlation
coefficient $C$ (close to 1), this does not necessarily imply that they are 
mechanistically related. They may be false correlations (FC), as opposed
to true correlations (TC), the latter signalling a true causal connection.
The matter lends itself to hilarious observations: the rate of drowning by falling in a pool 
appears tightly correlated with Nicolas Cage's movies: unless one assumes that Cage's movies are so
badly received to induce some to drown themselves, there is little question that this is a false correlation.
This case is trivial, but the general problem is not: distinguishing between TC's and FC's is an 
art, as the problem is both hard and important.
The embarrassing fact is that false correlations grow much more rapidly with 
size of data set under investigation than the true ones (the nuggets).
As recently proven by Calude and Longo \cite{LONGO} the TC/FC ratio is a very 
steeply decreasing function of data size. Meng, on the other hand, has shown that to be able to make statistically reliable inferences one needs to have access to a very substantial (i.e. $>50\%$) fraction of the data on which to perform one's machine learning \cite{MENG}.

Once again, how big is big enough to make reliable machine learning prediction remains a very open question.
To be sure, we are a very far cry from the comfortable inverse square root law of Gaussian statistics. What is clearly required in the field of big data and machine learning is many more theorems that reliably specify the domain of validity of the methods and the amounts of data to produce statistically reliable conclusions. One recent paper that sets out the way forward is by Karbalayghareh {\it et al.} \cite{KARBA}

\section{Life in a finite world: too much data is like no data}

\begin{figure}
\centering
\includegraphics[width=\textwidth]{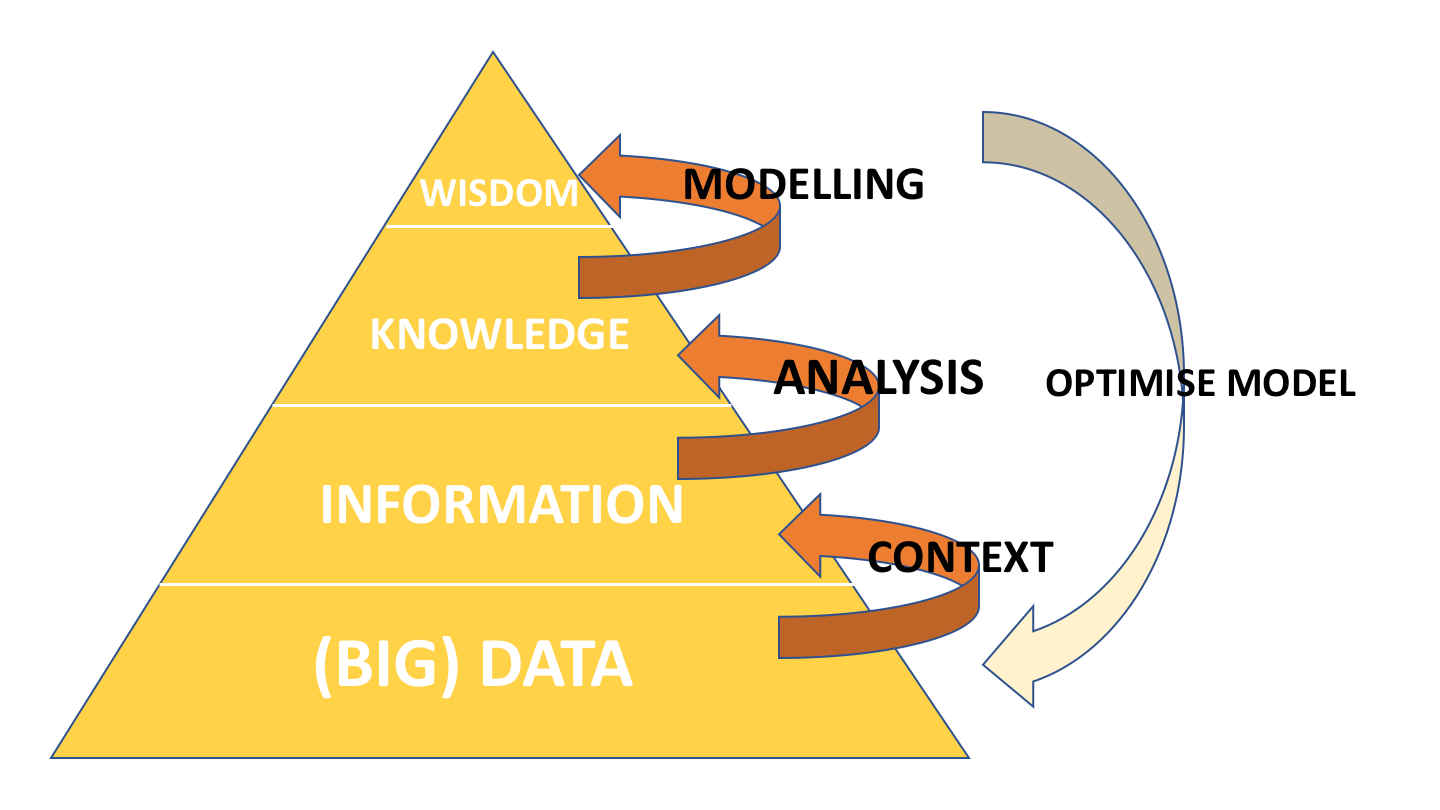}
\caption{A depiction of the DIKW pyramid to show the cooperation between big-data and modelling. It displays how ``data'', when put in context, leads to ``information''; analysing the ``information'' yields ``knowledge''; ``knowledge'' gained can be deeply understood by hypothesising a model for its underlying cause leading to ``wisdom'' which can be utilised to optimise the model by repeating the process.}
\label{fig:DIKW}
\end{figure}

Finally, there is an argument that hinges directly on epistemology and society, since it has to
do with a most prized human attribute: wisdom, the ability to take the right decision.
Wisdom is often represented as the top level of a pyramid of four, the
DIKW (Data-Information-Knowledge-Wisdom) chain, the one enabling us to take well-informed decisions.
From Data we extract Information, from Information we extract Knowledge
and finally from Knowledge we distil the ultimate goal: Wisdom, the ability to do the ``right thing''; see Figure \ref{fig:DIKW}.

Big Data driven decision theory is obviously of paramount importance to science, business and society, as
it is to each of our private lives. But, as a matter of fact, the ``constitutive relation'' between Data and Information, Information 
vs Knowledge and Knowledge vs Wisdom is not well known, to put it mildly. 
In the following, we shall argue that the pyramid representation is deceptive, for it conveys 
the idea that the layers stand in a simple linear relationship to one another, which is by no means the case. More importantly, it suggests that by expanding the basis (data) all upper-lying 
layers will expand accordingly, whence the mantra: more data, more wisdom. 

This flies in the face of a very general fact of life: sooner or later, all finite systems hit their ceiling.
the technical name of the game being {\it nonlinear saturation}, another well known concept in the
science of complex systems. 
This is the very general competition-driven phenomenon by which increasing 
data supply leads to saturation and sometimes even loss of information;
adding further data actually destroys information.
But let's discuss saturation first.
A well-known example of non-linear saturation is logistic growth in population dynamics.
Let $x$ be the number of individuals of a given species which reproduce at a rate $a>0$, say $a$ births per year.
In differential terms $dx/dt=ax$, leading to untamed exponential growth.
But obviously in a finite environment, with finite space and a finite amount of food, such untamed growth cannot last forever
for the environmental finiteness will necessarily generate competition, hence a depletion term.
Assuming competition is only between two individuals at a time, this results in the famous logistic equation
\begin{equation}
\label{LOGIS}
\frac{dx}{dt} = ax-bx^2
\end{equation}
where $b$ measures the strength of competition.
The right hand side is the epitome of what we mean, the rate of change growing linearly with $x$ but beyond  a certain 
threshold, $x^*=a/2b$, it decreases until it comes to a halt at $x=a/b$. 
By then the population stops growing, and the number of individuals left at that 
point is $x=a/b$, also known as the {\it capacity} of the system.
It is readily seen that the capacity goes inversely with the competition rate: the fiercer 
the competitors, the higher their needs and the lesser their number. 
As expected, big consumers present a threat, as is well known to those not driving 
four-by-four vehicles in the crowded streets of Rome and London.
 
The right hand side of (\ref{LOGIS}) can be cast in a more informative form as follows:
\begin{equation}
\label{LOGISTIC}
R(x) = bx(c-x)
\end{equation}
where $R$ indicates the effective rate and $c=a/b$ is the capacity.

This shows a nice symmetry (duality) between the population $x$ and the 
co-population $\bar x = c-x$, namely the gap between the actual number 
of individuals and the system's capacity.
Such symmetry is further exposed by writing the equations in dual form
\begin{eqnarray}
\frac{dx}{dt}                =  b x \bar x\\
\frac{d \bar x}{dt}        = -b x \bar x
\end{eqnarray}
Writing the equation in this manner highlights the dual process of generating population (``matter'') 
and annihilating co-population (``co-matter''). In passing, we note that the above system is invariant
under the exchange $x \leftrightarrow \bar x$, in combination with time inversion $ t \rightarrow -t$, which means that
the backward-time evolution of the co-population is the same as the forward-time evolution of the population. 
Such types of dual relations are typical of finite-size systems hosting nonlinear
cooperative/competitive interactions, in the generalised form
$$
R(x) = bx^{\alpha} \; (c-x)^{\beta}
$$    
where the exponents $\alpha$ and $\beta$, as well as the coefficients, may change depending 
on the specific phenomenon at hand. 
But the dual structure remains because it reflects the existence of a finite capacity.
The above example refers to population growth in time, which is not necessarily related to
the information growth with data.  
Nevertheless, it is our everyday experience that, beyond a certain threshold, further data does not add any 
information, simply because additional data contain less and less new information, and ultimately no new information at all.

This is quite common in complex systems: for instance, the number of degrees
of freedom of a turbulent flow (Information) grows like  $R^{9/4}$, where $R$ is the so-called Reynolds number
a dimensionless group measuring the strength of nonlinearity of fluid equations, whereas the volume 
of space (data) hosting the turbulent flow grows like $R^3$ (because $R$ scales like the linear size of the volume).
Hence the {\it information density}, i.e. the physical information per unit volume, scales like $I/V=R^{9/4-3}=R^{-3/4}$
a very steep decay at increasing Reynolds number. Given that  $R$ of the order of a million or more is 
a commonplace in real-life, it is clear that adding volume provides increasingly less 
return on investment in terms of gain of physical information.
We speculate, without proof, that this a general rule in the natural world. 

Let us now come to the worst-case scenario: data which destroy information. Eventually, additional data may even {\it contradict} previous data, perhaps because of inaccuracy but more
devious scenarios are not hard to imagine, thereby {\it destroying} information, because the new and the old data annihilate each other.  
In the latter scenario, information gain turns into information loss: seeing too much starts to
be like not seeing enough, to borrow from C.S. Lewis again.
We argue that such a dual trend applies to the DIKW chain as well, at least to the two lower layers:
{\it too much data is just like no data all}.
In fact this is possibly still more general: in a finite world, close to capacity, competitive 
interactions arise which either annihilate the return on investment (information per data unit)
or even make it negative, thereby destroying information and productivity,
over-communication being a well-known example in point\footnote{  
Incidentally, the reader should appreciate that away from capacity, the logistic equation (\ref{LOGISTIC}) 
equation  is linear because when $x \ll c$, then $c-x \sim c$}. 

Of course one can argue that, in actual practice, this depends on where the capacity is, so that 
BD can move it upwards and shift the problem away.
We argue that this threshold shifting is close to a chimera with lack of insight.
Unless we can apprehend the logical structures underlying any given phenomenon, we may just keep generating data conflicts that 
data accumulation alone will not be able to resolve, in fact quite the opposite.

In science, we strive to go from data-starved to data-rich, yet a blind data-driven 
procedure, as often advocated by the most enthusiastic BD neophytes, may well take us from 
data-rich to data-buried science, unless a just dose of theoretical reasoning is used 
as an antidote \cite{PVC2}.

More importantly, we all live in a finite world, and sooner or later its finite
capacity is going to be noticed. Even though such a basic reality is a taboo for many commercially inspired promoters of big data, we had better prepare for that.
Aggressive BD distracts attention from our limits, which is pretty dangerous 
for society long before it is for science.
Indeed, whatever philosophical stance one may adopt, it is clear 
that treating our resources as if they were unlimited is a sure way to 
disasters of various sorts, be they environmental, financial or social.
   
\section{Knowledge for Business: Big Data and Big Lies}

Good business is beneficial, and so is any healthy driver of economic wealth.
Hailing the pursuit of knowledge and innovation as a Trojan horse for business is not.

Big Data has a potentially enormous bearing on society, as is reflected by Pentland's recent book {\it Social Physics} \cite{PENTA}, the discipline which endeavours
to explain, predict and {\it influence} human behaviour (for good) based on physical-mathematical
principles, treating individuals as ``thinking molecules''.
The concept is technically appealing, although one clearly walking on a very thin tightrope between
good-willed science and social manipulation.
In principle, one can think of society as the ``material made up of thinking molecules'' and ask
how to best design such ``material'', so as to optimise moral values, while being bereft of cheating with flourishing 
economies, social equity and so on along a rosy-carpeted avenue.
In fact, such models provide scientific underpinning to Bauman's illuminating metaphor of
a liquid society as one where the rules change faster than most individuals can adjust to, 
leaving the majority behind \cite{BAU}.
Influencing human behaviour through ``healthy'' social pressure and flow of ideas is a noble goal, albeit 
one which walks on a high wire, the name of the (bad) game being plain manipulation for profit.
More precisely, what is hailed as pursuit of knowledge and innovation 
is in fact a very different goal, named {\it zero sales resistance}.

Zero sales resistance is an obvious goal for the most rapacious forms of capitalism, spinning around the money for money's sake paradigm,
instead of more conventional capitalism, in which money is the just and natural follow-on from healthy innovation, filling true societal gaps.

The damage done by BD brainware at the service of rapacious capitalism is all too evident: acquisition of private data 
in return for the dream of ``celebrity for everyone'' is a super-clever strategy, one that hits at the very roots of human 
weakness. 
  
Pentland, being well aware of the danger, invokes a new deal on data. But best intentions are easily fooled; so, here again C.S. Lewis appears apposite: ``When man proclaims conquest of power of nature, what it really means is conquest of power of {\it some} men over other men''.
When social media hit at human weaknesses, such as the desperate need for fame through
a growing list of ``followers'', collecting money for the disaster brought about by a tsunami might be a very good thing, but does 
not change the final balance sheet: mankind looses anyway.

A similar story applies to the big claims that cross the border into big lies, such as the promises of the so called ``Master Algorithm'',
allegedly capable of extracting {\it all} the information from the data, doing {\it everything}, just everything
we want, even before we ask for it. \cite{COVER}

In this essay, we hope we have made it clear why some of the boldest claims of
BD are in fact little or nothing short of big lies, that is:

\begin{enumerate}
\item Complex systems support uncertainty to a much stronger degree than the Law of Large Numbers
(Gaussian statistics) would have us believe. The implication is that error decay with data volume is considerably
slower, to the point of becoming impractically slow even in the face of zettabytes;

\item  No system is infinite, but when operating to their maximal extent, complex systems support the onset
of competitive interactions, in turn leading to data conflicts, which may either saturate the return on investment (in terms of the information gained
per unit of data) or even make it negative by supplying more data than a finite-capacity system can process.
Those BD aficionados who promise us  ``all we want and more'' simply choose to ignore this, and
it is not hard to see why.     

\item In the end, most of BD builds down to more or less sophisticated forms of curve fitting based on error
minimisation. Such minimisation procedures fare well if the error landscape is smooth, but they exhibit fragility towards corrugated ones in other situations, which are the rule in complex systems (Figure \ref{fig:landscapes}).    
\end{enumerate}      

Given these properties of nonlinear systems, the idea of replacing understanding with glorified curve fitting, no matter how ``clever'', 
appears a pretty questionable bargain, to put it mildly.
In this context, it is amusing to recall a conversation between 
Enrico Fermi and Freeman Dyson in 1952 \cite{FERMI}:\\\\
Fermi: How many parameters did you use in your calculations?\\
Dyson: Four.\\
Fermi: My friend John von Neumann used to say with four parameters, I 
can fit an elephant and with five, I can make him wiggle his trunk.\\

And with that, the conversation was over.
 
\section{What can be done?}

Once radical empiricism, hype-blinded high-tech optimism and the most rapacious forms of
business motivation, are filtered out,   what remains of Big Data is nonetheless a serious and promising scientific methodology.
In the end, however, it is nothing other than an elaborate form of curve fitting, but this is not intended as
a dismissive statement: sophisticated forms of inference, search and optimisation are involved in 
this activity, which deserve credit and respect, although certainly not awe.

There is no doubt that the ``big data/machine learning/artificial intelligence'' approach has plenty of scope to play a creative and important role in addressing major scientific problems. Among the applications, pattern recognition is particularly powerful in detecting patterns which might otherwise remain hidden indefinitely (modulo the problem of false positives mentioned earlier).  Possibly the most important role is likely to be in establishing patterns which then demand further explanation, where scientific theories are required to make sense of what is discovered. We have written elsewhere \cite{PVC1} of the fact that rapid ``successes'' of BD approaches take far longer to turn into sources of scientific insight.

In passing, however, we cannot refrain from commenting on the resurgence of use of the term ``artificial intelligence'' in this context, more than forty years after Marvin Minsky's unfortunate claim that computers were just a few years away from emulating human intelligence. That wild claim led to decades long ``AI winter'' from which one observes not only a thaw but the extravagant  hype accompanying any claimed successes of the BD approach.  We do not propose to digress into a discussion of AI, other than to point out that the concept has been subjected to penetrating analysis among others by Roger Penrose \cite{PENROSE}, who argues cogently that no digital computer will ever be capable of matching the human brain in terms of its ability to resolve problems such as those that reside in the class of the G\"{o}delian undecidable.  It matters not one iota that a so-called ``AI machine'' has the capability of assimilating the contents of staggeringly vast numbers texts (Tegmark \cite{TEG}).

In the final part of this essay, we focus rather on the positive aspects, namely how BD might assist with the struggle of the human mind to overcome three notorious barriers: non-linearity, non-locality and hyperdimensional spaces.

\subsection{Nonlinearity}
Nonlinearity is a notoriously tough cookie for theoretical modelling, for various
reasons, primarily because nonlinear systems do not respond in proportion to 
the extent to which they are prompted.
The most spectacular and popular metaphor of nonlinearity is the well known 
``butterfly'' effect, namely the little butterfly beating her wings in Cuba and triggering 
a hurricane in Miami in the process.
This is the ominous side of nonlinearity, the one that hits straight at our ability to predict
the future, the harbinger of uncertainty.

Less widely known perhaps is the sunny side of nonlinearity, that is its {\it constructive} power, which is
most apparent in biology where it underlies spatio-temporal organisation.
We shall not delve any further into this vital edifice of modern science \cite{FRONTIERS}.
Up to half a century ago, nonlinearity was hidden under the carpet of science for two good reasons. First, 
many systems under comparatively small loads, do respond linearly indeed (consider, for example, the logistic equation away from capacity). Second, linear systems are incomparably easier to be dealt with on mathematical grounds.
It was only in the 1960s, with the birth of chaos theory, that nonlinearity started to be fully embraced by
the scientific method, and it has continued to advance across all fields of science ever since.

While BD can certainly be of assistance in tackling some of the vagaries of non-linear systems, the fractal nature of many nonlinear dynamical systems utterly defies any notion of the smooth mappings upon which essentially all machine learning algorithms are based, rendering them nugatory from the outset in such contexts.  Indeed, the discontinuous nature of many nonlinear systems is simply not amenable to approaches based on machine learning's common assumptions that relationships are smooth and differentiable. 

\subsection{Nonlocality}

A further source of difficulty for scientific investigation is nonlocality understood as meaning the presence of long range correlations, by which we mean that interactions between entities (such as particles or fluid domains) decay very slowly with increasing distance. 
The classical example is the gravitational many-body problem, in which the force decays with the square of the inverse distance between two bodies. 
Nonlocality is a problem because it generates an all-to-all interaction scenario in which the
computational complexity grows quadratically with the number of interacting units.
The problem is far more acute in the quantum context, where nonlocality takes the form of the once-dreaded
``action at a distance'',  or more precisely to entanglement, meaning that different parts of a
system remain causally connected even when they are arbitrarily far apart.
This challenges our basic intuition that things interact most when they are in close proximity.

Leaving aside abundant metaphysical and science fiction based ramifications, entanglement stands as a 
highly counter-intuitive and difficult phenomenon to deal with by the current methods of theoretical science. Addressing quantum correlations with machine learning is plainly a major challenge.

\subsection{Hyper-dimensions}
We are used to living in three-spatial dimensions, plus time, and we are clearly often in difficulty in going further.
In fact, visualizing objects, not to mention dynamic phenomena, in just three dimensions seems to be 
complicated enough, as everyone dealing with visualisation software knows all too well.
Yet most problems inhabit a much larger domain, known as {\it phase-space}. 

Macroscopic systems consist of a huge number of individual components, typically in the 
order of the Avogadro number $A_v \sim 6 \times 10^{23}$ for the case of standard quantities of matter that we encounter. 
If each component is endowed with just six degrees of freedom, say its position in space
and its velocity, this makes six times Avogadro's number of variables, namely a mathematical
problem in six times Avogadro's number of dimensions.
Such is the monster-dimensionally that matters for many modelling purposes.

Thanks to a gracious  gift ``we neither understand, nor deserve'' in Eugene Wigner's words \cite{WIGNER},
much can be learned about these systems by solving problems in a much 
lower number of dimensions using the methods of statistical mechanics \cite{ARROW}. Even so, the task of modelling complex systems, say weather forecasting,
protein folding, just to name two outstanding problems in modern science,
remains very hard \cite{PVC1}.

Calculating the electronic structure of molecules is firmly in the class of computationally intractable problems. Accurate calculations scale factorially in the size of the basis sets used and render the highest levels of theory/accuracy essentially unattainable for anything other than the smallest of molecular systems. Here, considerable hype if not expectation has been focused on the construction of working quantum computers which would exhibit a special form of ``quantum parallelism'' that allows even such kinds of classically intractable problems to be solved on feasible time scales.  There is absolutely no chance of BD/ML/AI being applicable here: each problem is in a class of its own, and there are not going to be sufficient examples of solved problems available any time soon on which inference based approaches could even begin to be contemplated. 

What is curious about the current fad for quantum computing is that, as with BD etc., the hype is at its peak in the big corporations, such as Microsoft, Google, IBM, and so on who make claims we will have a working quantum computer in five years (we are excluding the D-Wave adiabatic quantum variant). These are the very same corporations which inundate us with reminders of the power of BD/ML/AI. And remarkably, one of the applications they say would be a ``killer app'' is in quantum chemistry, for the discovery of future drugs, at the same time as they promote BD methods to do the same thing. 

\section{A new scientific deal} 
It would be highly desirable if BD and particularly machine-learning techniques could
help surmount the three basic barriers to our understanding described above.

For now, however, in hard-core physical science at least, there is little evidence of any major BD-driven breakthroughs, at least 
not in fields where insight and understanding rather than zero sales resistance is the prime target: physics and chemistry do not succumb readily to the seduction of BD/ML/AI. It is extremely rare for specialists in these domains to simply  go out and collect vast quantities of data, bereft of any guiding theory as to why it should be done. There are some exceptions, perhaps the most intriguing of which is astronomy, where sky scanning telescopes scrape up vast quantities of data for which machine learning has proved to be a powerful way of processing it and suggesting interpretations of recorded measurements. In subjects where the level of theoretical understanding is deep, it is deemed aberrant to ignore it all and resort to collecting data in a blind manner. Yet this is precisely what is advocated in the less theoretically well grounded disciplines of biology and medicine, not to speak of social sciences and economics. The oft-repeated mantra of the life sciences, as the pursuit of ``hypothesis driven research'', has been cast aside in favour of large data collection activities \cite{PVC1}.

And, if the best minds are employed in large corporations to work out how to persuade people to click on online advertisements instead of cracking 
hard-core science problems, not much can be expected to change in the years to come. An even more acute story goes for social sciences and certainly for business, where  
the burgeoning growth of BD, more often than not fuelled by bombastic claims, is a compelling fact, 
with job offers towering over the job market to an astonishing extent. But, as we hope we have made clear in this essay, BD is by no means the panacea 
its extreme aficiodonas want to  portray to us and, most importantly, to funding agencies.
It is neither Archimedes' fulcrum, nor the end of insight.

Therefore, instead of rendering theory, modelling and simulation obsolete, BD should and will ultimately be used to
complement and enhance it. Examples are flourishing in the current literature, with machine 
learning techniques being embedded to assist large-scale simulations of complex systems
in materials science, turbulence \cite{KAXIR,JULIA,RUI} and also to provide major strides towards personalised
medicine \cite{PVC2}, a prototypical problem for which statistical knowledge will never be a replacement 
for patient-specific modelling \cite{PVC1}.  
It is not hard to predict that major progress may result from an inventive blend of the two, perhaps emerging as a 
new scientific methodology.

\section*{Acknowledgements}
SS wishes to acknowledge financial support  
from the European Research Council under the European
Union's Horizon 2020 Framework Programme (No. FP/2014-
2020)/ERC Grant Agreement No. 739964 (COPMAT). PVC is grateful for funding from the MRC Medical Bioinformatics project (MR/L016311/1), EU H2020 CompBioMed and VECMA (Grant Nos. 675451 and 800925) and from the UCL Provost. 

This essay grew out of the Lectio Magistralis ``Big Data Science: the End of the Scientific Method as We Know It?''
given by SS at the University of Bologna and various talks by PVC on the need of big theory for big data.
SS appreciates enlightening discussions with S. Strogatz and G. Parisi. PVC thanks E. Dougherty, F. Alexander and R. Highfield for valuable discussions.

\bibliography{bigdata}

\begin{thebibliography}{10}
\bibitem{COVER}
This sentence appears in the marketing introduction (Italian version) of the
  book, not in the book itself.
\end{thebibliography}

\end{document}